\RequirePackage{ifpdf}
\pdfoutput=1
\documentclass{JINST}
\usepackage{amsmath}

\title{Dual-Phase Liquid Xenon Detectors \\ for Dark Matter Searches}

\author{Marc Schumann \\
\llap{}Albert Einstein Center for Fundamental Physics, Universit\"{a}t Bern, CH-3012 Bern, Switzerland\\
  E-mail: \email{marc.schumann@lhep.unibe.ch}}

\abstract{Dual-phase time projection chambers (TPCs) filled with the liquid noble gas xenon (LXe) are currently the most sensitive detectors searching for interactions of WIMP dark matter in a laboratory-based experiment. This is achieved by combining a large, monolithic dark matter target of a very low background with the capability to localize the interaction vertex in three dimensions, allowing for target fiducialization and multiple-scatter rejection. The background in dual-phase LXe TPCs is further reduced by the simultaneous measurement of the scintillation and ionization signal from a particle interaction, which is used to distinguish signal from background signatures. This article reviews the principle of dual-phase LXe TPCs, and provides an overview about running as well as future experimental efforts. }


\begin{document}

\section{Dark Matter and Dark Matter Detection}

The standard model of cosmology, the $\Lambda$CDM model with only a handful of parameters, can very well describe all cosmological observations, the most prominent being the angular power spectrum of the cosmic microwave background (CMB). It is based on the idea that the observed accelerated expansion of the Universe is due to the effect of a cosmological constant $\Lambda$ (dark energy), contributing about 70\% to the Universe's energy density. The remaining fraction is mainly made up from cold dark matter (CDM, 25\%) which builds large scale structures via gravitational interactions. The dark matter particle is yet undetected. All known particles of the standard model of particle physics, mainly protons, neutrons and electrons (which form nuclei, atoms and molecules and hence all ``visible'' cosmic structures), as well as $\gamma$-rays and neutrinos only make up the remaining 5\% in the energy budget. The most precise numbers, as recently measured by the Planck satellite, are $\Omega_\Lambda=68.3$\%, $\Omega_\text{CDM}=26.8$\%, and $\Omega_\text{bary}=4.9$\%~\cite{ref::planck}.

The existence of dark matter has been proven indirectly by many astronomical and cosmological measurements at all length scales: from the rotation curves of spiral galaxies, via the dynamics of galaxy clusters, to the minute temperature fluctuations in the CMB spectrum, see~\cite{ref::dmevidence}. From the understanding of how structure has grown in the early Universe, as well as from detailed simulations of these processes, one can conclude that the dark matter must be cold, moving with non-relativistic velocities. Particles which have been thermally produced in the early Universe must therefore have a sizeable mass, which excludes neutrinos as dark matter candidates. Alternatively, the dark matter particles could have been produced in a phase transition, which is the case for axions~\cite{ref::sikivie}. Here, we will concentrate on thermally produced WIMPs (weakly interacting massive particles)~\cite{ref::wimp}, which are expected to have masses above $\sim$1\,GeV/$c^2$, and which interact very weakly with standard model particles. Such WIMPs arise naturally in various theories beyond the standard model, the most prominent being supersymmetry (WIMP=neutralino)~\cite{ref::susy} and large extra dimensions (WIMP=LKP)~\cite{ref::extradim}.

There are three different avenues pursued to experimentally detect the dark matter particle~ $\chi$~\cite{ref::dm2103}: production at colliders ($q\overline{q} \to \chi \overline{\chi}+X$), indirect detection of annihilation products ($\chi \overline{\chi} \to q\overline{q}, \ell \overline{\ell}, \ldots$), and direct detection of WIMP interactions~\cite{ref::directdet,ref::dmrev}. 
The small but finite interaction with standard model particles makes it in principle possible to detect WIMPs via their scattering off a nucleus in sensitive detectors ($\chi + N \to \chi + N +E_\text{rec}$). Interactions with atomic electrons are mass-suppressed as WIMPs are electrically neutral. In the most simple case -- coherent spin-independent (scalar) WIMP-nucleon scattering -- the cross section $\sigma_{SI}$ is proportional to the target's atomic mass-number squared, $A^2$, which favors heavy target materials. The expected interaction rate is very small and can be estimated by 
\begin{equation}
R \propto N^* \ \frac{\rho_\chi}{m_\chi} \ \langle \sigma_{SI} \rangle \lesssim 1\ \textnormal{evt/kg/y.} 
\end{equation}
$N^*$ is the number of target nuclei in the detector, $\rho_\chi \approx 0.3$\,GeV/cm$^3$ is the local dark matter density, $m_\chi$ the WIMP mass, and $\langle \sigma_{SI} \rangle$ is the velocity averaged scattering cross section. The approximate numerical value comes from current exclusion limits and has a WIMP mass dependence. The expected energy spectrum of the recoiling nucleus, the signature which is eventually measured in a detector, is a feature-less falling exponential with a mean energy of a ${\cal O}$(10)\,keV only.

Not only these small recoil energies and interaction rates make it difficult to detect WIMP dark matter, one also needs to take into account the abundant background from radioactive isotopes in the environment and the detector itself (mainly from the $^{238}$U and $^{232}$Th chains and $^{40}$K), producing $\alpha$, $\beta$, and $\gamma$-radiation, as well as neutrons via ($\alpha$,n) and spontaneous fission reactions. Cosmic rays also lead to background, the most dangerous being muon-induced neutrons, since the neutral WIMPs will -- similar to neutrons -- only interact with the nucleus (nuclear recoil), while all other backgrounds interact electromagnetically with the atomic electrons (electronic recoil). Taking into account all these boundary conditions, an ideal detector for the direct detection of WIMPs should feature
\begin{itemize}
 \item a large target mass of an isotope with a high mass number $A$,
 \item a low energy threshold,
 \item an ultra-low radioactive background, and
 \item a possibility to distinguish signal (nuclear recoils) from background (electronic recoils).
\end{itemize}
In order to reduce the cosmic-ray induced background, dark matter detectors have to be installed in deep-underground laboratories in tunnels or mines, where the overburden completely eliminates the hadronic component of the cosmic rays, and reduces the muon flux by 5-7\,orders of magnitude.

We will show in the next two Sections, that WIMP detectors based on dual phase time projection chambers, filled with a liquid xenon (LXe) target, fulfill all these requirements. Sections~4 and~5 present the status of a few ongoing and future experimental projects.

\section{Liquid Xenon as Dark Matter Target}

The noble gas xenon ($Z=54$, $\overline{A}=131.29$) is usually used as detection medium in a dark matter experiment in liquid form, cooled down to about $-95^\circ$C at an operation pressure of $\sim$2\,bar. When a particle interacts with the liquid xenon (LXe) target, the recoiling electrons (from interactions of $\beta$- and $\gamma$-rays) or nuclei (from neutrons and WIMPs) excite and ionize the xenon atoms. Some of the deposited energy is also transferred into atomic motion, however, since the LXe is not in crystalline form, this energy escapes detection. The excited Xe$^*$~atoms form excimer states Xe$^*_2$ with neutral xenon atoms, which subsequently decay under the emission of scintillation light in the vacuum-ultraviolet (VUV), at a mean wavelength of 178\,nm~\cite{ref::xewavelength}:
\begin{equation}\label{eq::scint}
\textnormal{Xe}^*  \xrightarrow{+\textnormal{Xe}} \textnormal{Xe}^*_2 \rightarrow 2\, \textnormal{Xe} + h \nu.
\end{equation}
The target is ionized as well and the ions form singly ionized molecules Xe$^+_2$ with neutral atoms. If the ionization electrons are not removed from the interaction site, they can recombine, leaving behind excited xenon atoms, which eventually will again decay via the emission of scintillation light: 
\begin{equation}\label{eq::ion}
\textnormal{Xe}^+ + e^- \xrightarrow{+\textnormal{2 Xe}} \textnormal{Xe}^+_2 + \textnormal{Xe} \xrightarrow{+e^-} 2\,\textnormal{Xe} + \textnormal{Xe}^{**} \to 2\,\textnormal{Xe} + \textnormal{Xe}^* + \textnormal{heat} \xrightarrow{\textnormal{(\ref{eq::scint})}} 4\, \textnormal{Xe} + h \nu.
\end{equation}

Eqs.~(\ref{eq::scint}) and~(\ref{eq::ion}) show three things: First, if one finds a possibility to extract the ionization electrons, not only the scintillation light but also the ionization charge signal from the interaction can be measured, provided that there is a charge sensitive detector. Second, if the electrons are removed, for example by an electric field, recombination is suppressed and the total light signal from the interaction is reduced. Third, the total number of quanta, photons $h\nu$ and electrons $e^-$, is constant for a given energy deposition (ignoring fluctuations and the energy which goes into heat). This means that the individual numbers are anti-correlated, which is exploited to improve the energy resolution of LXe detectors~\cite{ref::anticorrelation}. More details are discussed in~\cite{ref::chepel}.

\section{Dual Phase Time Projection Chambers}

\begin{figure}[b!]
\centering
\includegraphics[width=0.8\columnwidth]{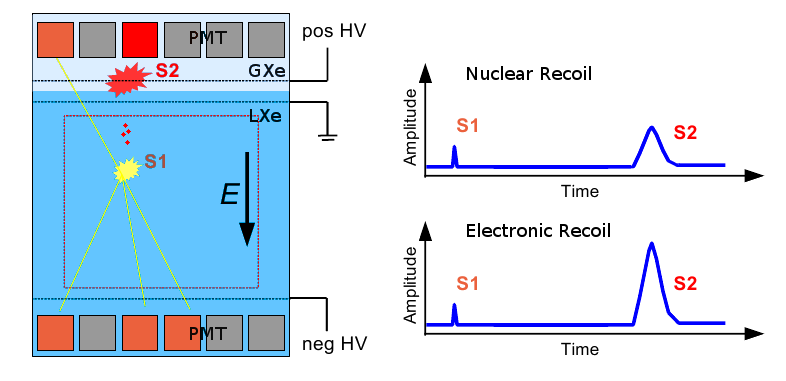}
\caption{A dual phase time projection chamber measures scintillation light (S1) and the ionization charge signal, which is converted to a proportional scintillation signal (S2) in the xenon gas phase (GXe). The time distance between the two signals and the pattern on the top PMT array is used to reconstruct the event vertex. The ratio S2/S1 is different for nuclear and electronic recoils and used for background discrimination. }
\label{fig::dualphase}
\end{figure}

The principle of a dual phase time projection chamber (TPC) filled with a noble gas (xenon or argon) has been first suggested by~\cite{ref::dolgoshein}. It was further developed and adapted for dark matter searches with LXe by the ZEPLIN-II~\cite{ref::zep2}, ZEPLIN-III~\cite{ref::zep3}, XENON10~\cite{ref::xe10}, XENON100~\cite{ref::xeInstr}, and LUX~\cite{ref::luxinstr} collaborations. The working principle is shown in Fig.~\ref{fig::dualphase}: A particle interaction in the LXe generates prompt scintillation light, see Eqs.\,(\ref{eq::scint}) and (\ref{eq::ion}), and ionizes the noble gas. An electric field across the target volume removes the ionization electrons from the interaction site by drifting them upwards to the xenon gas phase (GXe) above the liquid. The electrons are extracted into the GXe by a second, stronger electric field where they again excite the xenon atoms and generate a secondary scintillation light signal (S2), which is proportional to the amount of charge liberated by the interaction. The S2 signal, as well as the primary scintillation light (S1) are observed by two arrays of photomultipliers (PMTs), installed below the target in the LXe and above in the gas.

The simultaneous detection of light and charge allows for the 3-dimensional reconstruction of the event vertex in the TPC and provides several means to reject radioactive backgrounds:
\begin{itemize}
 \item Xenon is a rather dense material ($\rho \approx 3$~g/cm$^3$) and absorbs $\gamma$-radiation efficiently. Backgrounds from the experiment's surroundings and the detector construction materials will hence interact mainly in the outer parts of the detector, leaving the inner region with a considerably reduced background level. Only this inner target with its low background is used for the WIMP search (``fiducialization'').
 \item The expected WIMP cross section is so small that it is virtually impossible that WIMPs interact more than once in the detector. The charge measurement, however, allows for the identification and rejection of multiple scatter events, which exhibit more than one S2 signal.
 \item The energy loss d$E$/d$x$ in the LXe is different for electronic recoils (ER) from a $\gamma$- or $\beta$-background interactions and for nuclear recoils (NR) from a potential WIMP signal (or a neutron). The energy partition into excitation and ionization varies for the two types of recoils, leading to a different ratio of the charge-to-light ratio S2/S1. This ratio is used to further reject backgrounds. $\alpha$-particles produce a huge amount of light and constitute no relevant background in a dual phase TPC.
\end{itemize}
\smallskip

\begin{figure}[tb]
\centering
\includegraphics[width=0.8\columnwidth]{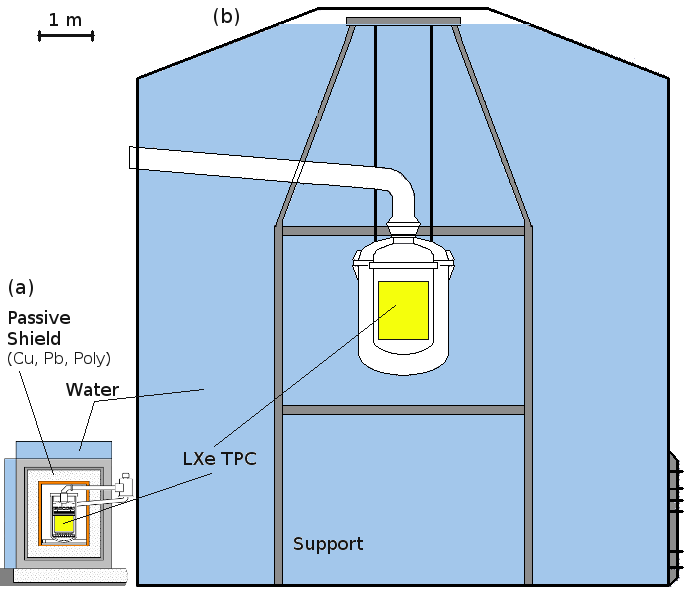}
\caption{Comparison of a future LXe dark matter experiment to a presently running one using the same scale. The drawing is based on the designs of (a) XENON100~\cite{ref::xeInstr} and (b) XENON1T~\cite{ref::xe1t}, which is currently under construction and is expected to take data by 2015. While the smaller system, with a LXe inventory of 161\,kg and a target mass of 62\,kg, can still be installed in a massive shield made from water, lead, copper, and polyethylene, the much larger new detector ($\sim$3000\,kg total, $\sim$2200\,kg target) will be placed in the center of a large water shield, which is operated as a \u{C}erenkov muon veto. }
\label{fig::xenon}
\end{figure}

One of the most efficient ways to reduce background is to avoid it in the first place. Dark matter detectors are therefore placed deep underground (against charged cosmic rays and muons) installed inside massive shields made from low-$Z$ (against neutrons) and high-$Z$ materials (against $\gamma$-radiation) or several meters of water, see Fig.\,\ref{fig::xenon}. The detectors are constructed from specially selected radio-pure materials (see for example~\cite{ref::xe100_screening}) and the design is made such that all known ``hot'' detector components are placed outside of the shield.

\section{Recent Results}

The results from the experiments XENON10~\cite{ref::xe10}, ZEPLIN-II~\cite{ref::zep2}, and ZEPLIN-III~\cite{ref::zep3} as well as several years of R\&D prepared the way to the dual-phase LXe detectors XENON100~\cite{ref::xeInstr} and LUX~\cite{ref::luxinstr}, which are leading the field of direct WIMP detection in terms of sensitivity at the time being. In the following, we focus on recent results of these two projects.

\subsection{XENON100}

\begin{figure}[tb]
\centering
\includegraphics[width=0.8\columnwidth]{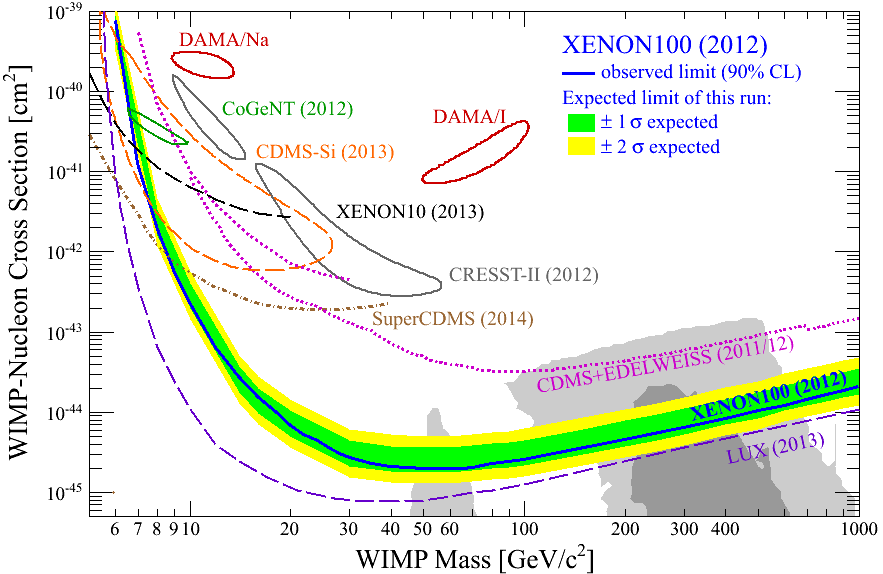}
\caption{Current experimental results on spin-independent WIMP-nucleon interactions. The signatures, which could be interpreted as being due to WIMP interactions, observed by DAMA/LIBRA~\cite{ref::dama}, CoGeNT~\cite{ref::cogent}, CRESST-II~\cite{ref::cresst}, and CDMS-Si~\cite{ref::cdms_si} are strongly challenged by the null results from other experiments, in particular from the dual-phase LXe TPCs XENON100~\cite{ref::xe100_run10} and LUX~\cite{ref::lux}, as well as from the low-mass WIMP result by XENON10~\cite{ref::xe10s2only} and SuperCDMS~\cite{ref::supercdms}. Figure adapted from~\cite{ref::xe100_run10}.}
\label{fig::limit}
\end{figure}

XENON100~\cite{ref::xeInstr} is a dual-phase LXe TPC, installed in a low-background stainless steel cryostat, which resides inside a passive shield made from water, lead, polyethylene, and copper (Fig.\,\ref{fig::xenon}, a) at the Italian Laboratori Nazionali del Gran Sasso (LNGS). The cylindrical active target mass of 62\,kg (30\,cm height and diameter) is viewed by two arrays of 1''$\times$1'' PMTs (Hamamatsu R8520), which are optimized to operate in LXe. Apart from a few interruptions for maintenance, XENON100 is operating stably since 2009 and published the most sensitive results on spin-independent WIMP-nucleon scattering, for WIMP masses above $\gtrsim$8\,GeV/$c^2$, in 2010, 2011, and 2012~\cite{ref::xe100_run7,ref::xe100_run8,ref::xe100_run10}. No indication for an excess of events above the expected background has been observed in any of these results, leading to strong upper limits. As shown in Fig.~\ref{fig::limit}, these null-results severely challenge the interpretation of the results from DAMA/LIBRA~\cite{ref::dama}, CoGeNT~\cite{ref::cogent}, CRESST-II~\cite{ref::cresst}, and CDMS-Si~\cite{ref::cdms_si} as being due to WIMP interactions. With the latest science run published in 2012~\cite{ref::xe100_run10}, based on a total exposure of 225\,live days $\times$ 34\,kg, XENON100 has reached its design sensitivity of $\sigma_\text{SI}=2\times10^{-45}$\,cm$^2$ at $m_\chi\sim50$\,GeV/$c^2$. The data has also been interpreted in terms of spin-{\it de}pendent WIMP-nucleon interactions, leading to some of the most stringent upper limits to date~\cite{ref::xe100_run10sd}.

The amplification in the gas phase, converting the ionization electrons into proportional scintillation light, is very efficient and allows the detection of single electrons with dual phase TPCs~\cite{ref::singlee_zep,ref::singlee_xe}. This has been exploited in the previous detector XENON10, where an analysis using only the charge signal led to very strong limits at low WIMP masses, at the price of an increased background due to the lack of discrimination and largely reduced fiducialization power~\cite{ref::xe10s2only} (see Fig.~\ref{fig::limit}). A recent publication of the XENON100 collaboration studied the response of the detector to single electron charge signals in detail~\cite{ref::singlee_xe}: one electron creates secondary photons, which lead to a typical detected light signal of $\text{S2}\sim20$\,photoelectrons. The low energy S2~spectrum is clearly quantized, with overlapping distributions from 1, 2, 3, $\ldots$~electron signals. These are mainly caused by photo-ionization of surfaces and impurities, leading to a correlation of their rate with large S2~signals (the photon source) and with the LXe purities (the photo-ionized target). This high statistics sample of very small signals allows the monitoring of the detector stability, as well as the measurement of instrument parameters such as the charge extraction yield.

Dedicated scattering experiments have been performed in order to measure the response of LXe to nuclear recoils of energy $E_\text{nr}$, given in keV$_\text{nr}$ (nuclear recoil equivalent)~\cite{ref::manzur,ref::plante}. The signal quenching is usually expressed as relative scintillation efficiency ${\cal L}_\text{eff}(E_\text{nr})=\text{LY}(E_\text{nr})/\text{LY}(^{57}\text{Co})$, the light induced by a nuclear recoil of energy $E_\text{nr}$ normalized by the light yield of a 122\,keV $\gamma$-ray from a $^{57}$Co source. The parametrizations of ${\cal L}_\text{eff}(E_\text{nr})$ used by XENON and LUX are almost identical, and the validity of the choice has been recently confirmed by a XENON100 study comparing the low-energy response of the detector to an $^{241}$AmBe neutron source with a detailed Monte Carlo simulation~\cite{ref::ambe_mc}: After taking into account efficiencies, energy resolution and the $^{241}$AmBe source strength, the spectra agree very well down to 3\,keV$_\text{nr}$, which is well below the XENON100 analysis threshold of 6.6\,keV$_\text{nr}$. The relative scintillation efficiency ${\cal L}_\text{eff}(E_\text{nr})$ and the charge yield $Q_y(E_\text{nr})$ derived from this study are in good agreement with the direct measurements (with $Q_y$ showing a somewhat slower increase below 8\,keV$_\text{nr}$ than~\cite{ref::manzur}).

\begin{figure}[tb]
\centering
\includegraphics[width=0.9\columnwidth]{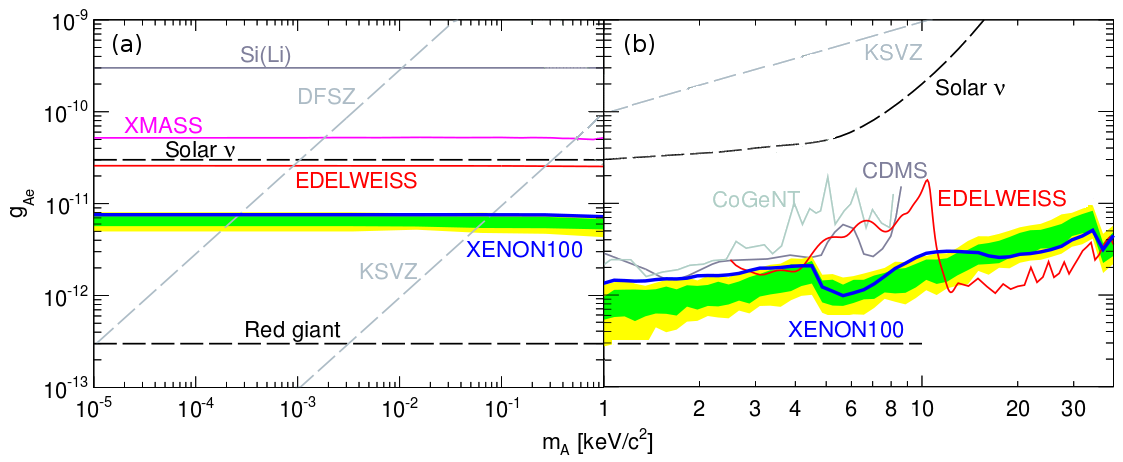}
\caption{Upper limits from XENON100 on (a) solar axions and (b) galactic axion-like particles (ALPs), where one assumes that the entire amount of dark matter is made up from ALPs. These particles are expected to interact with the LXe via the axio-electric effect~\cite{ref::axioelectric} with coupling constant $g_{Ae}$, leading to electronic recoil signatures. The expected WIMP signal, however, is nuclear recoils. Figure adapted from~\cite{ref::axions}.}
\label{fig::axions}
\end{figure}

Similar scattering measurements were also performed to establish a S1 signal-based low energy scale for electronic recoils~\cite{ref::keVee_cu,ref::keVee_uzh}. The results largely agree and with the expectations of a sophisticated model~\cite{ref::nest}, allowing the XENON100 collaboration to place limits on the coupling constant $g_{Ae}$ of solar axions and galactic axion-like particles (a hypothetical light-mass particle which could make up the entire amount of dark matter)~\cite{ref::ringwald}. These are expected to interact with the LXe target via the axio-electric effect~\cite{ref::axioelectric}, leading to observable electronic recoil signatures. The low background of the XENON100 detector led to some of the best limits on these particles so far~\cite{ref::axions}, as shown in Fig.\,\ref{fig::axions}.

\subsection{LUX}

The LUX experiment~\cite{ref::luxinstr}, installed in the SURF underground facility in South Dakota (USA), is currently the largest and most sensitive dark matter detector in operation. Its dual-phase TPC of 48\,cm height and 47\,cm diameter encloses 250\,kg of LXe (out of 370\,kg total mass) and is installed in a low-background titanium cryostat, which itself is placed inside a water shield of 7.6\,m diameter and 6.1\,m height. Prompt light (S1) and proportional scintillation signal (S2) are detected using Hamamatsu R8778 photomultipliers with a diameter of 2``. 

Due to its large mass (leading to improved self-shielding) and high light collection efficiency (leading to a very low threshold of 3\,keV$_\text{nr}$) LUX was able to publish very strong first results in 2013, based on an exposure of of 85\,live days $\times$ 118\,kg. No excess of events above the expected background was observed, leading to the currently most stringent exclusion limits, with a minimum of $\sigma_\text{SI}=7.6\times10^{-46}$\,cm$^2$ at $m_\chi=33$\,GeV/$c^2$ (90\% CL)~\cite{ref::lux}, see Fig.~\ref{fig::limit}. The LUX result fully confirms the XENON100 constraints at low WIMP masses and  further increases the tension with the ''hints`` for low mass WIMPs. The design goal of LUX is to reach cross sections down to $2.0 \times 10^{-46}$\,cm$^2$ after running for 300\,live days.

\section{The Future}

While LUX is expected to reach its design goal by 2015~\cite{ref::luxtalk}, other dual-phase LXe experiments will come online soon. The first phase of the Chinese experiment PandaX~\cite{ref::panda}, featuring a flat, pancake-shaped target of 120\,kg LXe ($\sim$25\,kg fiducial mass), is currently running in the Jinping laboratory (China) and first results are expected in the next months. 

At LNGS (Italy), the XENON collaboration is constructing XENON1T~\cite{ref::xe1t}, a new dual-phase LXe TPC with a target mass of 2.2\,tons ($\sim$3.0\,tons total), which will be the first TPC with a WIMP search mass of about 1\,ton. The detector will be installed in a large water shield, operated as \u{C}erenkov muon veto, see Fig.~\ref{fig::xenon} (b). The design goal of XENON1T is to perform a background-free dark matter run with an exposure of 1\,ton $\times$ 2\,years, improving the current XENON100 sensitivity by 2\,orders of magnitude in order to reach $\sigma_\text{SI}=2 \times 10^{-47}$\,cm$^2$ at $m_\chi\sim50$\,GeV/c$^2$. XENON1T is fully funded, construction is ongoing, and data taking is expected to commence in 2015.

About 2~order of magnitude beyond the science reach of XENON1T, direct WIMP searches will be eventually limited by irreducible neutrino backgrounds~\cite{ref::irreducible}. However, as the interaction cross section is not constrained by theory~\cite{ref::strege} and as one would like to measure the WIMP-induced recoil spectrum with high statistics, in case WIMP dark matter will be found in the next-generation experiments, the community already thinks about even larger experiments exploring the remaining parameter space. Basically all sub-systems of XENON1T are being constructed such that they can be re-used in a later upgrade phase of the experiment, XENONnT, with about 7\,tons of LXe. This will dramatically reduce the cost and the time needed compared to the development of a a new experiment. The LUX collaboration is proposing the similarly-sized LZ detector~\cite{ref::lz}.

The ultimate WIMP detector, covering basically all accessible parameter space above the limit from coherent neutrino-nucleus scattering, is studied within the DARWIN project~\cite{ref::darwin}: Spin-independent cross sections down to the few $10^{-49}$cm$^2$ level around $m_\chi \sim 50$\,GeV/$c^2$ can be reached with TPCs using a LXe target of 20\,tons ($\sim$14\,tons fiducial mass). Such large LXe detectors can not only search for WIMPs, but can also measure the flux of solar pp-neutrinos in real time and are sensitive to the neutrinoless double beta decay of $^{136}$Xe, even without isotopic enrichment~\cite{ref::darwinnu}.


\end{document}